\documentstyle[prd,aps,floats]{revtex}
\begin{document}
\draft
%
%
\input epsf
\renewcommand{\topfraction}{0.8}
\twocolumn[\hsize\textwidth\columnwidth\hsize\csname
@twocolumnfalse\endcsname

\preprint{CERN-TH/96-302, SUSSEX-AST 96/10-2, astro-ph/9610183}
\title{Complete power spectrum for an induced gravity open inflation model}
\author{Juan Garc\'{\i}a-Bellido}
\address{Astronomy Centre, University of Sussex, Falmer, 
Brighton BN1 9QH, United Kingdom\\ and 
Theory Division, CERN, CH-1211 Geneva 23, Switzerland}
\author{Andrew R.~Liddle}
\address{Astronomy Centre, University of Sussex, Falmer, 
Brighton BN1 9QH, United Kingdom}
\date{\today}
\maketitle
\begin{abstract}
We study the phenomenological constraints on a recently proposed
model of open inflation in the context of induced gravity. The main
interest of this model is the relatively small number of parameters,
which may be constrained by many different types of observation. We
evaluate the complete spectrum of density perturbations, which
contains continuum sub-curvature modes, a discrete super curvature
mode, and a mode associated with fluctuations in the bubble wall.
>From these, we compute the angular power spectrum of temperature
fluctuations in the microwave background, and derive bounds on the
parameters of the model so that the predicted spectrum is compatible
with the observed anisotropy of the microwave background and with
large-scale structure observations. We analyze the matter era and
the approach of the model to general relativity. The model passes
all existing constraints.
\end{abstract}
\pacs{PACS numbers: 98.80.Cq \hspace*{2mm} 
Preprint CERN-TH/96-302, SUSSEX-AST 96/10-2, astro-ph/9610183}
 
\vskip2pc]

\section{Introduction}

The inflationary paradigm~\cite{book} not only provides a solution to
the classical problems of the hot big bang cosmology, but also
predicts an almost scale invariant spectrum of metric perturbations
which could be responsible for the observed anisotropy of the cosmic
microwave background (CMB), as well as the origin of the large-scale
structure. Present microwave background anisotropy experiments offer
only weak constraints; for example the COBE satellite \cite{COBE}
gives a very accurate determination of the amplitude of large-angle
anisotropies (which can be used to normalize theories) but only weakly
constrains the shape of the spectrum. Information is beginning to come
in on degree scales, and from combining microwave anisotropy
constraints with those from large-scale structure, but the present
situation still offers considerable freedom. However, we can expect
this to change dramatically in the near future, especially with the
launch of new-generation microwave anisotropy satellites {\em MAP}
\cite{MAP} and {\em COBRAS/SAMBA} \cite{COBSAM} which promise to
measure both cosmological parameters such as $\Omega_0$, $H_0$ and
$\Omega_{{\rm B}}$, and parameters associated with the primordial
spectra to great accuracy~\cite{JKKS}. It is therefore desirable to
provide a variety of inflationary models with definite predictions,
which could be used to test and exclude them.

Until recently, inflation was always associated with a flat
universe, due to its ability to drive the spatial curvature so
effectively to zero.  However, it is now understood that inflation
comprises a wider class of models, some of which may give rise to an
open universe at present~\cite{open,BGT,LM}. Observations suggesting a
high value of the Hubble parameter, such as those using the Hubble Space
Telescope~\cite{Freed}, have motivated the idea of
considering a low-density universe, in an attempt to make the age
compatible with globular cluster ages. Most frequently a cosmological
constant is introduced to restore spatial flatness, but open inflation
models (see Ref.~\cite{intro} for an introduction) offer the
alternative of a genuinely open universe. Such models generically
contain a field trapped in a false vacuum which tunnels to its true
vacuum via nucleation of a single bubble, inside which a second period
of inflation drives the universe to almost flatness. This way one
solves the homogeneity problem independently from the flatness
problem, allowing for an open homogeneous universe inside the bubble.

In an open universe, the analysis of density perturbations and
microwave anisotropies is considerably more complicated than in the
usual flat space case. Early studies by Lyth and Stewart \cite{LySt}
and by Ratra and Peebles \cite{RaPe} evaluated the spectrum for
slow-roll models leading to an open Universe, using the conformal
vacuum as an initial condition. In the single-bubble models a
different vacuum choice is appropriate, leading to a slightly
different spectrum \cite{sasaki,BGT}. It was later realised however
that extra perturbations, with discrete wavenumbers, can also be
generated. In all, three different types of perturbation have been
identified: a continuous spectrum of modes with wavenumber $k$ greater
than the curvature scale, known as sub-curvature modes; a
super-curvature mode associated with the open de Sitter vacuum
\cite{sasaki,LW}, and a mode associated with perturbations in the
bubble wall at tunneling~\cite{bubble,Jaume,YST}. The observed
large-scale structures are due to the sub-curvature modes, but
large-angle microwave anisotropies are generated by all three types,
with observations seeing the combined total anisotropy. The first
computation of all three types of mode together for a particular model
was made in Ref.~\cite{bubble} for arbitrary $\Omega_0$ in the context
of the two-field models of Linde and Mezhlumian~\cite{LM}. Later on, a
thorough calculation of all three contributions from the point of view
of quantum field theory in open de Sitter space was carried out by
Yamamoto et al.~\cite{YST} (see also Ref.~\cite{super}). Here we shall
carry out a similar calculation for a different two-field model, for
which we shall also discuss some of the implications of large-scale
structure observations.

In addition to scalar metric perturbations, we expect open inflation
to lead to the production of a gravitational wave spectrum, as in
conventional inflationary models. Unfortunately, no-one has yet
formulated a method of calculating this spectrum even approximately,
and so we shall not be able to consider them here. In chaotic
inflation models, gravitational waves are negligible in the slow-roll
limit (see e.g.~Ref.~\cite{LL}); one can hope that this is also
true in the open inflation case, but that remains to be confirmed.

The particular open inflation model we shall study, introduced in
Ref.~\cite{GL}, is based on the induced gravity Lagrangian \cite{Zee}.
The interest of this model is the relatively small number of
parameters, which can be constrained by several different types of
observation.  The inflaton is a dilaton field, whose vacuum
expectation value at the end of inflation determines the present value
of the gravitational constant. We will constrain the model from CMB
and large-scale structure observations, as well as ensuring that
post-Newtonian and oscillating gravitational coupling bounds are
satisfied.

\section{The Model}

We consider the model of Ref.~\cite{GL}, with an induced gravity Lagrangian 
\cite{Zee}
\begin{equation}
\label{induced}
{\cal L} = {1\over2}\xi\varphi^2R - {1\over2} \partial_\mu\varphi
        \partial^\mu\varphi + V(\varphi) + {\cal L}_{\rm mat} \,.
\end{equation}
The dilaton field $\varphi$ determines the effective gravitational
coupling, which is positive for $\xi>0$. In the absence of a potential, this 
action corresponds to the usual Brans--Dicke action~\cite{BD}, where
\begin{equation}
\Phi = 8\pi\xi\,\varphi^2\ , \hspace{1cm} \omega={1\over4\xi}\,.
\end{equation}
The Einstein and scalar field equations are~\cite{Weinberg,GBL}
\begin{eqnarray}
- \xi \varphi^2 G_{\mu\nu} &=&  g_{\mu\nu} V(\varphi) + \xi
        (\nabla_\mu \nabla_\nu - g_{\mu\nu} \nabla^2) \varphi^2
        \nonumber \\[1mm]
&& + \Big(\partial_\mu \varphi \partial_\nu \varphi  - 
        {1\over2} g_{\mu\nu} (\partial \varphi)^2\Big) + T_{\mu\nu}\,,
\end{eqnarray}
and
\begin{equation}
\nabla^2 \varphi = - V'(\varphi) + \xi \varphi R \,,
\end{equation}
where $G_{\mu\nu}$ is the Einstein tensor. 
Using the identities $G_{\mu\nu}^{\hskip4mm;\nu} = 0$ and 
$R_{\mu\nu}\nabla^\nu\Phi = \nabla_\mu(\nabla^2\Phi) - \nabla^2
(\nabla_\mu\Phi)$, together with the $\varphi$ equation of motion,
we find that the energy-momentum tensor is conserved, 
\begin{equation}
\label{cons}
T_{\mu\nu}^{\hskip3mm;\nu} = 0\,,
\end{equation}
even in the presence of a potential for the scalar field. Substituting
$R$ into the equation of motion of the scalar field, we obtain
\begin{equation}
\label{phieq}
{1\over2}(1+6\xi)\nabla^2 \varphi^2 = 4V(\varphi) - \varphi
        V'(\varphi) + T^\lambda_{\ \ \lambda} \,.
\end{equation}

We will consider a potential of the type~\cite{Zee,GL}
\begin{equation}
\label{potential}
V(\varphi) = {\lambda\over8} (\varphi^2-\nu^2)^2\,.
\end{equation}
During a radiation dominated era $T^\lambda_{\ \ \lambda} = 0$ and
the scalar field will sit at its minimum; matching the present-day Planck 
mass demands
\begin{equation}
\label{Planck}
8\pi\,\xi\nu^2 = m_{{\rm Pl}}^2 \,.
\end{equation}
During a matter era, the scalar field oscillates around its
minimum with a large frequency and negligible amplitude, passing 
all the tests associated with an oscillating gravitational 
coupling~\cite{Will} as we will show in Section~\ref{matter}.

\section{Induced gravity open inflation}

In the induced gravity open inflation model \cite{GL}, the initial
period of inflation is driven by the false vacuum energy of a second
scalar field $\sigma$.  This energy density is able to hold the
dilaton at a fixed location displaced from the minimum of its
potential. After the false vacuum decays, the rolling of the dilaton
to its minimum drives the second period of inflation necessary to give
$\Omega_0$ in the desired range.

\subsection{False vacuum inflation}

Initially the scalar field $\sigma$ is in its false vacuum. The details of 
its potential are not particularly important; we will parametrize them 
later. In the false vacuum, the universe expands driving the spatial 
curvature and any previous inhomogeneities to zero. Later on, the $\sigma$ 
field tunnels to its true minimum at $V(\sigma) = 0$, via the production of 
a bubble. 

The equations of motion of the dilaton field before and after the tunneling 
can be written as~\cite{GL} 
\begin{eqnarray}
H^2 + 2H{\dot\varphi\over\varphi} + {K\over a^2} &=& {1\over3\xi\varphi^2}
        \left[{1\over2} \dot\varphi^2 + V(\varphi) + V(\sigma) 
        \right] \,, \\[1mm]
\ddot\varphi+3H\dot\varphi+{\dot\varphi^2\over\varphi} 
        &=& {4V(\varphi)-\varphi V'(\varphi) +4V(\sigma)\over(1+6\xi)
        \varphi} \,.
\end{eqnarray}
Here $V(\sigma)=V_0$ in the false vacuum and vanishes in the true vacuum, 
while the curvature $K$ is effectively zero before tunnelling and is 
negative afterwards. The basis for the open inflation model is the existence 
of a stable static solution in the false vacuum \cite{GL}, with
\begin{eqnarray}
\label{static}
\varphi_{{\rm st}}^2 &=& \nu^2\,\Big(1 + {8V_0\over\lambda\nu^4}\Big)
        \equiv \nu^2 \, (1+\alpha)\,,\\[1mm]
H_{{\rm st}}^2 &=& {8\pi V_0\over3m_{{\rm Pl}}^2} \,.
\end{eqnarray}
Its stability is best seen in the Einstein frame \cite{GL}. Under the 
transformation
\begin{eqnarray}
dt &=& {\nu\over\varphi} \, d\tilde t \,, \hspace{5mm} 
        a(t) = {\nu\over\varphi}\,\tilde a(\tilde t) \,, \\[1mm]
{\phi\over\nu} &=& (1+6\xi)^{1/2} \ln {\varphi\over\nu}\,,
\end{eqnarray}
the effective potential in the false vacuum becomes
\begin{eqnarray}
\label{UF}
U_{{\rm F}}(\phi) &=& {\lambda\nu^4\over8} \, \left[
        1-2{\nu^2\over\varphi^2} + (1+\alpha){\nu^4\over\varphi^4}
        \right]\,,\\[1mm]
U_{{\rm F}}'(\phi) &=& {\lambda\nu^3\over2(1+6\xi)^{1/2}}
        \, {\nu^2\over\varphi^2} \, \left[ 1 - (1+\alpha)
        {\nu^2\over\varphi^2} \right] \,,\\[1mm]
U_{{\rm F}}''(\phi) &=& {\lambda\nu^2\over1+6\xi} \,
        {\nu^2\over\varphi^2} \, \left[2(1+\alpha)
        {\nu^2\over\varphi^2}-1\right]\,,
\end{eqnarray}
where primes denote derivatives with respect to the Einstein-frame 
scalar field $\phi$. It is clear that $U_{{\rm F}}'(\phi_{{\rm st}})=0$ at 
the static value, while the effective square mass is positive ensuring 
stability. At the static point, we have
\begin{eqnarray}
\label{false}
H_{{\rm F}}^2 &=& {8\pi \,U_{{\rm F}}\over 3m_{{\rm Pl}}^2}
        = {\lambda\nu^2\over24\xi}\,{\alpha\over1+\alpha}\,,\\
\label{false2}
m_{{\rm F}}^2 &\equiv& U_{{\rm F}}''(\varphi_{{\rm st}}) =
        {\lambda\nu^2\over1+6\xi}\,{1\over1+\alpha}\,,
\end{eqnarray}
where $H_{{\rm F}}$ is the rate of expansion of the universe in the Einstein
frame and $m_{{\rm F}}$ is the mass of the $\phi$ field at the static point.

\subsection{True vacuum inflation}

Eventually the $\sigma$ field tunnels to its true vacuum by nucleating
a bubble, inside which the universe inflates to almost flatness. A 
sufficiently low tunneling rate ensures that the bubble stays isolated 
\cite{GS,GL}. After tunneling, the effective potential in the true vacuum, 
again in the Einstein frame, becomes
\begin{eqnarray}
\label{UT}
U_{{\rm T}}(\phi) &=& {\lambda\nu^4\over8}\,\left(1-{\nu^2\over\varphi^2}
        \right)^2\,,\\[1mm]
U_{{\rm T}}'(\phi) &=& {\lambda\nu^3\over2(1+6\xi)^{1/2}} \,
        {\nu^2\over\varphi^2} \, \left(1-{\nu^2\over\varphi^2}
        \right)\,,\\[1mm]
U_{{\rm T}}''(\phi) &=& {\lambda\nu^2\over1+6\xi} \, {\nu^2\over\varphi^2}
        \, \left(2{\nu^2\over\varphi^2}-1\right)\,.
\end{eqnarray}
The minimum of the potential is now at $\varphi = \nu < \varphi_{{\rm st}}$, 
and the field slow-rolls from $\varphi_{{\rm st}}$ driving a second stage 
of inflation. The dynamics of this situation were investigated long ago 
in Ref.~\cite{azt}. The rate of expansion and effective mass in the true 
vacuum immediately after tunneling are 
\begin{eqnarray}
\label{true}
H_{{\rm T}}^2 &=& {8\pi \,U_{{\rm T}}\over 3m_{{\rm Pl}}^2} -\frac{K}{a^2}
        = {\lambda\nu^2\over24\xi} \, \left({\alpha\over1+\alpha}
        \right)^2 - \frac{K}{a^2} \,,\\
m_{{\rm T}}^2 &\equiv& U_{{\rm T}}''(\varphi_{{\rm st}}) = 
        {\lambda\nu^2\over1+6\xi} \, {1-\alpha \over (1+\alpha)^2} \,.
\end{eqnarray}
The curvature term quickly becomes negligible as the second phase of 
inflation progresses. 

For later use, we also define the usual slow-roll 
parameters~\cite{LL} soon after tunneling
\begin{eqnarray}
\label{epet}
\epsilon &\equiv& \frac{m_{{\rm Pl}}^2}{16 \pi} \left( 
        \frac{U_{{\rm T}}'(\phi)}{U_{{\rm T}}(\phi)} \right)^2 =
        {8\xi\over1+6\xi}\,{1\over\alpha^2}\,,\\[1mm]
\label{epet2}
\eta &\equiv& \frac{m_{{\rm Pl}}^2}{8 \pi} 
        \frac{U_{{\rm T}}''(\phi)}{U_{{\rm T}}(\phi)} =
        {8\xi\over1+6\xi}\,{1-\alpha\over\alpha^2}\,.
\end{eqnarray}

The scalar field $\phi$ slow-rolls down the effective potential given by
Eq.~(\ref{UT}) until it starts oscillating around its minimum and inflation
ends. The value of $\phi$ at the end of inflation can be computed
from the condition \mbox{$-\dot H_{{\rm T}} \simeq H_{{\rm T}}^2$} (or 
equivalently $\dot\phi^2 \simeq U_{{\rm T}}(\phi)$), giving
\begin{equation}
\label{end}
\varphi_{{\rm end}}^2 = \nu^2\,\left(1 + {8\xi\over1+6\xi}\right) 
        \equiv \nu^2\,(1+\beta) \,.
\end{equation}
The number of $e$-folds during the second stage of inflation from
$\phi_{{\rm st}}$ to $\phi_{\rm end}$ can be computed in the Einstein frame,
\begin{eqnarray}
\label{efolds}
N &=& {1\over\xi\nu^2}\,\int_{\phi_{\rm end}}^{\phi_{{\rm st}}}
        {d\phi\,U_{{\rm T}}(\phi)\over U_{{\rm T}}'(\phi)} \nonumber \\
&=& {1\over\beta}\,\left[\alpha-\beta-\ln\left(1+\alpha\over
        1+\beta\right)\right]\,.
\end{eqnarray}
In order to produce an open universe, the number of $e$-folds after
tunneling has to be around $N = 60$, the precise number depending on
the reheating temperature and other details of the post-inflationary
evolution. We adopt the number 60 for definiteness. This gives a
relation between the two dimensionless parameters $\alpha$ and $\beta$.

\section{Metric perturbations and temperature anisotropies}

Quantum fluctuations of the inflaton field $\phi$ produce long-wavelength 
curvature perturbations; we will use ${\cal R}$ to denote the curvature 
perturbation on comoving hypersurfaces (in the Einstein frame). 

Open inflation generates three different types of modes: those that
cross outside during the second stage of inflation and constitute a
continuum of sub-curvature modes; a discrete super-curvature mode
associated with the open de Sitter vacuum, and a mode associated with
the bubble wall fluctuations at tunneling. The mode functions are
eigenvalues of the Laplacian, with eigenvalue $-k^2$ where $k$ is the
wavenumber. Defining $q^2 = k^2 -1$, then the sub-curvature
modes have positive $q^2$ and the other modes have negative $q^2$. We label 
the former mode functions $\Pi_{ql}(r)$ and the latter 
$\bar{\Pi}_{|q|l}(r)$. In Appendix A we give explicit forms for these; see 
also Ref.~\cite{modes}.

The spectrum ${\cal P}_{\cal R}(q)$ of the curvature perturbation can be 
defined from the mode expansion of ${\cal R}$ by \cite{LW}
\begin{equation}
\label{spectrum}
\langle{\cal R}_{qlm}  {\cal R}_{q'l'm'} \rangle = 
        \frac{2\pi^2}{q(q^2+1)} \, {\cal P}_{\cal R}(q) \, \delta(q-q')
        \delta_{ll'} \delta_{mm'}.
\end{equation}

In order to compare with observations, we must compute the effect that 
such a perturbation has on the temperature of the CMB, expanded 
as usual in spherical harmonics
\begin{equation}
\label{temp}
{\Delta T\over T}(\theta,\phi) = \sum_{lm} a_{lm} \, Y^l_m(\theta,\phi)\,.
\end{equation}
The main contribution on large scales comes from the Sachs--Wolfe 
effect~\cite{SW67}. The complete angular power spectrum 
$C_l\equiv\langle|a_{lm}|^2\rangle$ has contributions from the continuum of 
sub-curvature modes, the super-curvature mode and the bubble-wall mode,
\begin{equation}
\label{CL}
C_l = C_l^{(C)} + C_l^{(S)} + C_l^{(W)} \,.
\end{equation}

The contribution of each mode to the $C_l$ is measured by a window 
function $W_{ql}$, given by~\cite{SW67} 
\begin{equation}
\label{window}
5\,W_{ql} = \Pi_{ql}(\eta_0) + 6\int_0^{\eta_0} dr\,F'(\eta_0-r) \,
        \Pi_{ql}(r)\,,
\end{equation}
for the sub-curvature modes; the same expression with $\bar{\Pi}_{|q|l}$ 
gives the window function $\bar{W}_{|q|l}$ for the negative $q^2$ modes. 
Here
\begin{equation}
\label{F}
F(\eta)=5\,{\sinh^2\eta-3\eta\sinh\eta+4(\cosh\eta-1)\over
        (\cosh\eta-1)^3}\,,
\end{equation}
gives the growth rate of perturbations during the matter 
era~\cite{Mukhanov}, and $\eta_0 = \cosh^{-1} (2/\Omega_0-1)$ is the 
distance to the last scattering surface. The normalization of the 
contribution to the $C_l$ is given in the expressions below.

\subsection*{Sub-curvature modes}

A detailed computation of the amplitude of the sub-curvature modes gives the 
result~\cite{sasaki,BGT,YST}
\begin{equation}
\label{sub}
{\cal P}_{\cal R}(q) = \coth(\pi q) \, 
        \frac{8 \, U_{{\rm T}}}{3 \, \epsilon\, m_{{\rm Pl}}^4} \,,
\end{equation}
where $\epsilon$ is the slow-roll parameter defined earlier and we have 
dropped a small correction term from the change in mass during 
tunneling~\cite{YST}. The $\coth(\pi q)$ factor can be interpreted as due to 
the initial transient behavior as the curvature term dies away.

Notice that Eq.~(\ref{sub}) has only been derived in the case of perfect de 
Sitter expansion after tunneling~\cite{YST}. It seems very plausible that it 
also holds when there are deviations from de Sitter, where the right hand 
side is to be evaluated when $k=aH$. This is the only simple
formula which reduces to the correct result both for de Sitter expansion
and in the flat-space limit (see e.g.~Ref.~\cite{LLrep}) which must be 
attained after the curvature term has died away sufficiently.

Following normal practice, we describe the variation in the spectrum caused 
by the time variation of $H$ and $\epsilon$ by a power-law. Notice that this 
power-law is superimposed on the $\coth$ behavior, so the complete spectrum 
does not have a power-law form on very large scales. The power-law index of 
${\cal P}_{{\cal R}}(q)/\coth(\pi q)$ can be derived in the usual way from 
the slow-roll parameters as~\cite{LL} 
\begin{equation}
\label{tilt}
n -1 = - 6\epsilon + 2\eta = -\,{8\xi\over1+6\xi} \,
        {2(2+\alpha)\over\alpha^2} \,,
\end{equation}
and gives the standard result in the flat-space limit where the $\coth$ term 
in Eq.~(\ref{sub}) equals unity. This expression is valid provided both 
$\epsilon$ and $|\eta|$ are much less than one.

For later comparison, we write the spectrum as
\begin{equation}
{\cal P}_{\cal R}(q) = A_{{\rm C}}^2 \,\coth(\pi q) \,
        \left[ 1+q^2 \right]^{(n-1)/2} \,,
\end{equation}
where 
\begin{equation}
\label{RC}
A_{{\rm C}}^2 = \frac{8 \,U_{{\rm T}}(\phi_{{\rm
        st}})}{3\,\epsilon(\phi_{{\rm st}}) \, m_{{\rm Pl}}^4}
\end{equation}
is a measure of the amplitude at the $q = 0$ limit. Formally the 
spectrum diverges there, though not in a harmful way thanks to the window
function given by Eq.~(\ref{window}). 

For our model, Eq.~(\ref{RC}) becomes
\begin{equation}
\label{dens}
A_{{\rm C}}^2= {\lambda\over(16\pi\,\xi)^2} \,
        {1+6\xi\over6\xi} \left({\alpha^2\over1+\alpha}\right)^2\,.
\end{equation}

The angular power spectrum for the continuum modes can be written 
as~\cite{sasaki} 
\begin{equation} 
\label{powerC}
C_l^{(C)} = 2\pi^2\,\int_0^\infty
        {dq\over q(1+q^2)}\,{\cal P}_{\cal R}(q)\,W_{ql}^2\,.
\end{equation}
We compute the angular power spectrum for different values of $\Omega_0$ in 
the low-density range $0.2\leq\Omega_0\leq0.6$. In Fig.~1 we show the first 
twelve multipoles, adopting the notation $D_l = l(l+1)\,C_l$.\footnote{Our 
  calculation only includes the
  Sachs--Wolfe effect, as is appropriate for computing the amplitude
  on the largest angular scales. We therefore don't include the rise
  to the acoustic peak, caused by the first oscillation of the
  photon--baryon fluid, which is known to induce an effective extra
  tilt of around 0.15 \cite{BSW}; see for example Fig.~8 in Ref.~\cite{BW}
  (who only consider sub-curvature modes).}

\begin{figure}[t]
\centering 
\leavevmode\epsfysize=6cm \epsfbox{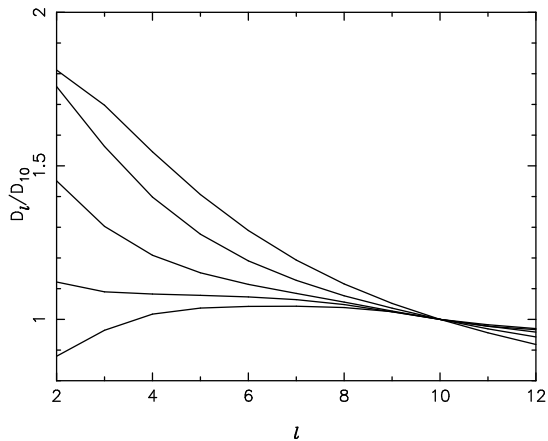}\\ 
\caption[fig1]{\label{fig1} The first 12 multipoles of the angular 
power spectrum associated with the continuum modes, normalized to the tenth 
multipole, for $\Omega_0 = 0.2, 0.3, 0.4, 0.5, 0.6$ as read from top to 
bottom at low $l$.}
\end{figure} 

The normalization to COBE for tilted open models has recently been given by 
Bunn and White \cite{BW}, under the assumption that only the continuum 
modes are important. They specify a quantity $\delta_{{\rm H}}$, which 
measures the normalization of the present matter power spectrum. The 
preferred value depends on $n$ and $\Omega_0$; however our model is nearly 
scale-invariant and the dependence on $\Omega_0$ is quite weak and can be 
ignored at the accuracy we are working. Therefore we take the value 
$\delta_{{\rm H}} = 2 \times 10^{-5}$ regardless of $\Omega_0$. In an open 
universe $\delta_{{\rm H}}$ is related to ${\cal P}_{\cal R}$ as \cite{BW}
\begin{equation}
\delta_{{\rm H}} = {2\over5} \, {\cal P}_{\cal R}^{1/2} 
        \, \frac{g(\Omega_0)}{\Omega_0} \,,
\end{equation}
where $g(\Omega_0)$ is a function measuring the suppression in the
growth perturbations relative to a critical-density universe, and
${\cal P}_{\cal R}$ is evaluated at around the 10th multipole where
$\coth (\pi q) \simeq 1$. The $\Omega_0$ dependence can give a factor of
up to 1.5 in the region of interest, but we can ignore it as we do not 
require such accuracy. Reproducing the amplitude of temperature anisotropies 
is the main constraint on the parameters of the model, and yields
\begin{equation}
\label{constrC}
\sqrt\lambda = 6 \times 10^{-3}\,\left({\xi^3\over1+6\xi}
        \right)^{1/2} {1+\alpha\over\alpha^2}\,,
\end{equation}
as found in Ref.~\cite{GL}. This relation can readily be satisfied for 
reasonable values of the parameters~\cite{GL}.

\subsection*{Super-curvature mode}

\begin{figure}[t]
\centering 
\leavevmode\epsfysize=6cm \epsfbox{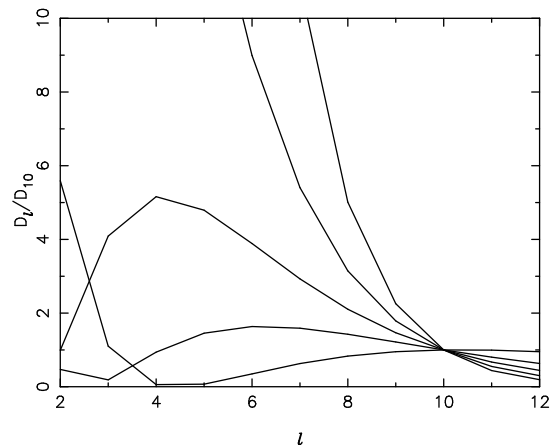}\\ 
\caption[fig2]{\label{fig2} As Fig.~1, but for the discrete 
supercurvature mode, showing $\Omega_0= 0.6, 0.5, 0.4, 0.3, 0.2$, reading 
from top to bottom at in the centre of the figure.} 
\end{figure} 

We now consider the contribution to the CMB aniso\-tropies coming from
the discrete super-curvature mode associated with the dilaton field $\phi$.
This mode appears in the open de Sitter spectrum when $m_{{\rm F}}^2<2
H_{{\rm F}}^2$ in the false vacuum~\cite{sasaki}. The tunneling field 
$\sigma$ does not have this mode in its spectrum, since its mass at the 
false vacuum should be much larger than the rate of expansion in order to
prevent tunneling via the Hawking-Moss instanton~\cite{LM}. The
wavenumber associated with this mode is given by
\begin{equation}
\label{super}
k^2 = 1 - \left[ \left({9\over4}-{m_{{\rm F}}^2\over H_{{\rm F}}^2}
        \right)^{1/2} - {1\over2}\right]^2\,.
\end{equation}
The amplitude of this mode is~\cite{super} 
\begin{equation}
\label{RS}
A^2_{{\rm S}} \simeq {8 \, U_{{\rm F}} \over 3\, \epsilon \,m_{\rm Pl}^4} =
        A^2_{{\rm C}} \, {U_{{\rm F}} \over U_{{\rm T}}} \,.
\end{equation}
where the normalization of $A^2_{{\rm S}}$ is defined through the formula 
for the angular power spectrum of temperature aniso\-tropies induced by this 
super-curvature mode, namely~\cite{YST,super} 
\begin{equation}
\label{powerS}
D_l^{(S)} \equiv l(l+1)\,C_l^{(S)} = 4\pi\, A_{{\rm S}}^2
        \, \bar{W}_{1l}^2\,.
\end{equation}
Fig.~2 shows the first twelve multipoles, showing quite a complicated
dependence. For example, it is not automatic that the quadrupole receives
the biggest contribution~\cite{GZ}. 

More important than the shape is the amplitude of these anisotropies
relative to the sub-curvature ones. We will compare their contributions in 
Section~\ref{comparison}.

\subsection*{Bubble wall mode}

In addition to the sub- and super-curvature modes, there is a
contribution from the bubble wall fluctuations. These fluctuations
contribute as a transverse traceless curvature perturbation mode with
$k^2=-3$, see Refs.~\cite{bubble,Jaume,YST}, which still behaves as
a homogeneous random field~\cite{Hamazaki,modes}.

Unlike the modes we've discussed so far, these modes need extra
parameters for their description, because their amplitude depends on
the details of the bubble wall, which is determined by the potential
for the $\sigma$ field.  This extra freedom allows the bubble wall
fluctuations to be tuned relative to the others.

The perturbation amplitude for the bubble wall mode is given
by~\cite{bubble,Joanne,YST}
\begin{equation}
\label{wall}
A^2_{{\rm W}} = {4\,U_{{\rm T}}\over  
        a^2b \, m_{\rm Pl}^4}\,\Big[a^2+(1+a^2b)^2\Big]^{1/2}\,,
\end{equation}
where 
\begin{equation}
\label{AB}
a^2 = {24\pi\,U_{{\rm T}}\,S_1^2\over m_{\rm Pl}^2\,
        (U_{{\rm F}}-U_{{\rm T}})^2}\,, \hspace{5mm}
b = {U_{{\rm F}}-U_{{\rm T}}\over4\,U_{{\rm T}}} \,.
\end{equation}
and $S_1$ gives the bubble wall contribution to the bounce action,
$B_{\rm wall} = 2\pi^2 R^3 S_1$ (see e.g.~Ref.~\cite{bubble}). In order
to compute $S_1$, we will consider a symmetry breaking potential of
the type
\begin{equation}
\label{US}
U(\sigma) = U_{{\rm F}} + {\gamma\over4}\,\sigma^2 (\sigma-\sigma_0)^2
        - \mu U_0 \Big({\sigma\over\sigma_0}\Big)^4\,,
\end{equation}
where $\sigma_0 = M\sqrt{2/\gamma}$ corresponds to the true vacuum and
$U_0=M^4/16\gamma$ is the value of potential at the maximum. With
$\mu\ll1$ for the thin-wall approximation to be valid, $S_1$ can be
computed as~\cite{bubble}
\begin{equation}
\label{S1}
S_1 = \int_0^{\sigma_0} d\sigma\,[2(U(\sigma)-U_{{\rm F}})]^{1/2} \simeq 
{M^3\over3\gamma}\,.
\end{equation}

\begin{figure}[t]
\centering 
\leavevmode\epsfysize=6cm \epsfbox{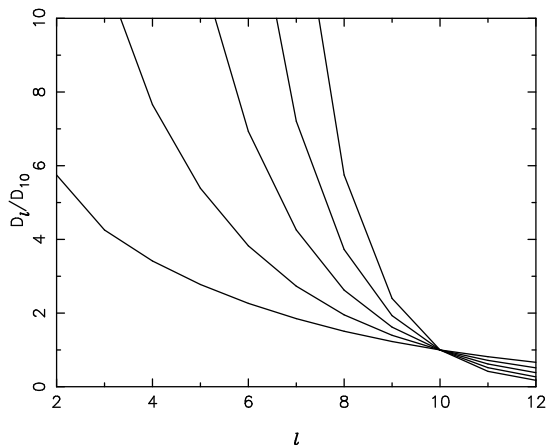}\\ 
\caption[fig3]{\label{fig3} As Fig.~1, but for the discrete bubble wall 
mode, for $\Omega_0= 0.6, 0.5, 0.4, 0.3, 0.2$, from top to bottom at low 
$l$.} 
\end{figure} 

In the limit $a^2b\ll1$ of small gravitational effects
at tunneling, we recover the result of Ref.~\cite{LM}, namely
\begin{equation}
\label{wall1}
A^2_{{\rm W}} = {2\,U_{{\rm T}} (U_{{\rm F}}-U_{{\rm T}}) \over \pi\,
     m_{\rm Pl}^2\,S_1^2} = A^2_{{\rm C}}\,{3\,\epsilon\over2\,a^2b} \,,
\end{equation}
where $\epsilon$ is the slow-roll parameter. However, in the opposite
limit of strong gravitational effects, $a^2b\gg1$, we have~\cite{bubble}
\begin{equation}
\label{wall2}
A^2_{{\rm W}} = {4\,U_{{\rm T}} \over m_{\rm Pl}^4} 
        = A^2_{{\rm C}} \, {3\,\epsilon\over2}\,,
\end{equation}
which is much smaller than the amplitude of the continuum modes.

The angular power spectrum associated with the bubble wall
mode is~\cite{bubble}
\begin{equation}
\label{powerW}
D_l^{(W)} \equiv l(l+1) \, C_l^{(W)} = {4\pi\,A^2_{{\rm W}}
        \over (l+2)(l-1)}\,\bar{W}_{2l}^2\,.
\end{equation}
Fig.~3 shows the first twelve multipoles. The quadrupole has the largest 
amplitude for all $\Omega_0$. 

We will compare their contribution to the CMB in the next Section.

\section{Comparison with observations}
\label{comparison}

\subsection{Microwave background anisotropies}

We can now examine constraints on the shape of the combined spectrum.
The COBE data alone do not offer particularly strong constraints in
this respect; for example, although Yamamoto and Bunn~\cite{YB} argued
that the inclusion of super-curvature modes could harm the fit to COBE
based on the two-year data, a recent comprehensive analysis of the
four-year data by G\'{o}rski et al.~\cite{GRSSB} finds no useful
constraint. Those papers however discussed only a particular model for
the super-curvature modes and didn't include the bubble-wall modes at
all. As Sasaki and Tanaka discussed~\cite{super}, there can be
interesting constraints if the super-curvature modes have their
amplitude enhanced, and we shall also se that the bubble wall modes
are typically more important than the super-curvature ones.

Unfortunately, the complicated structure of the perturbation spectra
in open inflation models means that for a full analysis each model
would have to be confronted with the COBE data set on a case-by-case
basis. Such an analysis is outside the scope of this paper. We
shall adopt a more simplistic approach, which is to demand that the
discrete modes do not dominate the quadrupole while contributing
negligibly to the tenth multipole. This would give an unacceptable shape to 
the spectrum on the scales sampled by COBE. 

\begin{figure}[t]
\centering 
\leavevmode\epsfysize=6cm \epsfbox{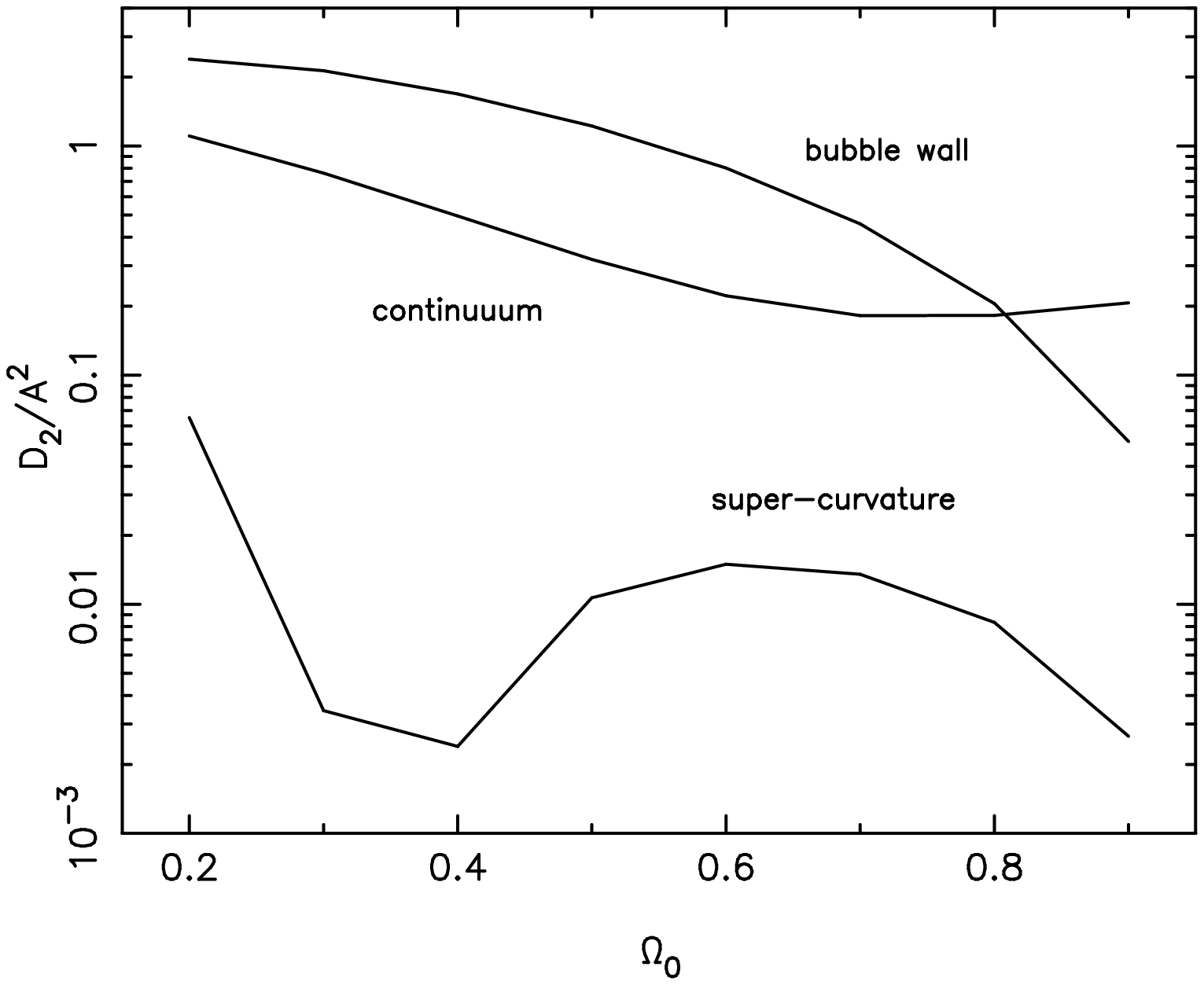}\\ 
\leavevmode\epsfysize=6cm \epsfbox{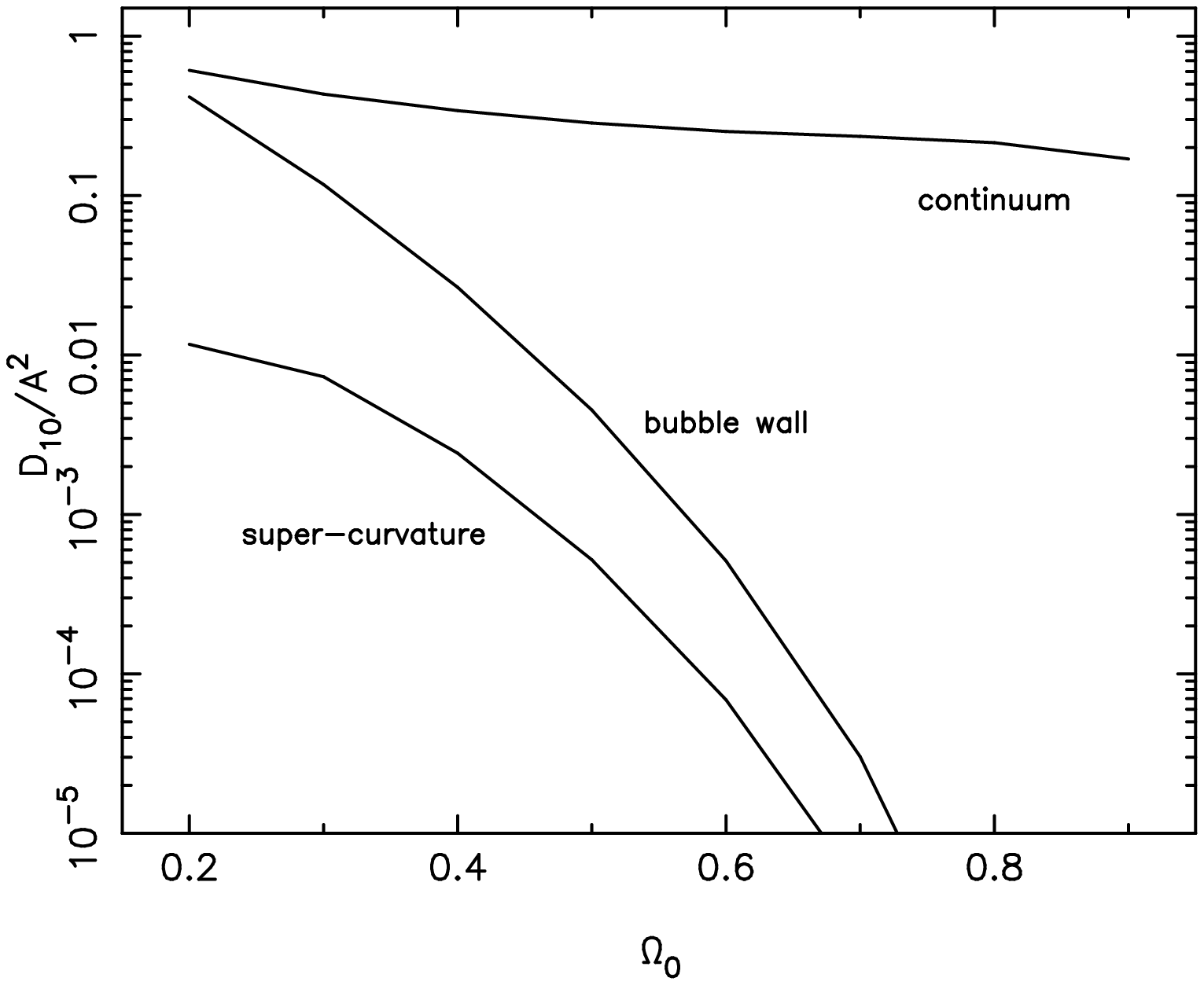}\\ 
\caption[fig4]{\label{fig4} The multipoles associated with each of the 
  modes, normalized to the 
  corresponding metric perturbation ($A_{{\rm C}}^2$, $A_{{\rm S}}^2$ or
  $A_{{\rm W}}^2$ as appropriate), as a function of $\Omega_0$. The top
  panel shows the quadrupole, while the lower shows the tenth multipole.}
\end{figure} 

In Fig.~4, we show the contributions to the quadrupole and to the tenth 
multipole, as functions of $\Omega_0$, normalized to the size of the 
corresponding metric perturbation. It does not require much effort to keep 
the contribution to the tenth multipole from the discrete modes low (unless 
$\Omega_0$ is very small), but we must ensure that the quadrupole is 
not dominated by the discrete modes.

Considering first the super-curvature modes, across the whole range of
interesting $\Omega_0$ we find that the super-curvature mode
contributes about a factor thirty less to the low multipoles than do
the continuum modes, for the same size of metric 
perturbation.\footnote{Although there is a dip around
$\Omega_0\simeq0.4$ caused by an `accidental' cancellation between
intrinsic and line-of sight terms, the dip is not at the same
location for the $l = 3,4...$ multipoles so the dip doesn't allow
one to weaken the constraint in its neighborhood~\cite{GZ}.} Unless
$\Omega_0$ is very small, then if the super-curvature mode
contribution is comparable to that of the continuum modes for the
quadrupole, then it is negligible at the tenth multipole.
Conservatively then, we shall require
\begin{equation}\label{condS}
A_{{\rm S}}^2 \lesssim 100 \, A_{{\rm C}}^2 \,,
\end{equation}
independently of $\Omega_0$, to prevent the super-curvature modes from 
dominating the low multipoles. In principle this limit could become 
inappropriate at small enough $\Omega_0$, because then the shape of the 
super-curvature spectrum is not so steep as to be ruled out by observations. 
However, by then the super-curvature modes are contributing substantially 
even to the tenth multipole, and this will force down the normalization ot 
COBE and make it impossible to fit large-scale structure observations. We 
can therefore adopt the above constraint even at low $\Omega_0$ values. 
This imposes only a very mild constraint on the parameters of the model, 
namely
\begin{equation}
\label{constrS}
\alpha > 0.01\,,
\end{equation}
which is easy to satisfy as we will see in Section~\ref{scenarios}. Since 
the value of $\Omega_0$ at present is not known with any reasonable
accuracy, it would be excessive to give a precise $\Omega_0$-dependent
constraint. However, in the future, with a narrow range of values for
$\Omega_0$, one would be able to give a more refined constraint on the
parameters of the model from a full comparison of the detailed
spectrum against COBE or its successors.

The same arguments can be applied to the bubble-wall modes; again we
must prevent the domination of the quadrupole by these modes. For the
interesting $\Omega_0$ values (i.e.~those not too close to one), we
see from Fig.~4 that this simply requires
\begin{equation}
\label{condW}
A_{{\rm W}}^2 \lesssim A_{{\rm C}}^2 \,,
\end{equation}
again independent of $\Omega_0$ in the interesting range. This again imposes 
only a very mild constraint on the parameters of the model.
For $a^2b\ll1$, we have $3 \epsilon < 2 a^2b$ giving
\begin{equation}
\label{constrW}
\gamma < 9\times 10^4\,{1+6\xi\over6\xi}\,
        {M^3\over m_{{\rm Pl}}^3}\,\alpha^{5/2}\,,
\end{equation}
which is easy to satisfy for sufficiently large $M$. For $a^2b\gg1$, the 
amplitude of the metric perturbation Eq.~(\ref{wall2}) is completely 
negligible since $\epsilon\ll1$, see Section~\ref{scenarios}.

\subsection{Implications for large-scale structure}

We can now combine the COBE normalization with observations of large-scale 
structure. This has the advantage of being relatively insensitive to the 
inclusion of super-curvature or bubble-wall modes, because with the 
constraints above these modes affect only the lowest
multipoles, and while these can influence whether or not the spectral
shape is a good fit to observations, they are not very significant at
the higher multipoles (around the tenth to fifteenth) which are most
important for determining the normalization. The drawback though is
that looking to large-scale structure introduces a dependence on all
the other cosmological parameters, namely the Hubble parameter $h$,
the baryon density $\Omega_{{\rm B}}$ and the nature of the dark
matter. Despite this, interesting constraints can still be found, and
two analyses have appeared which discuss tilted open models --- Liddle
et al.~\cite{LLRV} who used the two-year COBE data and, more recently,
White and Silk~\cite{WS} who used an accurate normalization to the
four-year COBE data. Both of these considered only cold dark matter;
other choices tend to strengthen the constraints so we shall do
likewise.

In the model we are considering, the spectral index is always tilted
to $n$ less than one, as seen from Eq.~(\ref{tilt}). Whether or not
this is allowed depends quite sensitively on both $\Omega_0$ and $h$.
If $\Omega_0$ is too low, below around $0.30$, then the open cold dark
matter model fares badly against observations; this conclusion is
consistent with a similar constraint from velocity flows \cite{D94}
which is independent of the power spectrum. For example, White and
Silk~\cite{WS} find that this value is allowed only if the power
spectrum is `blue', with $n$ at least $1.10$.  This conclusion is
enforced both by the cluster abundance and by the shape of the galaxy
correlation function. Our model will therefore be ruled out if it
turns out that the universe is indeed as open as this.

However, one does not have to increase $\Omega_0$ by very much to
radically change this conclusion. For $\Omega_0 = 0.5$, for example,
White and Silk find viable models for $n$ as low as $0.85$, with the
preferred value depending on the Hubble parameter $h$. Our model can
therefore be comfortably compatible with the data for this $\Omega_0$,
and at least marginally compatible for $\Omega_0$ as low as $0.4$.

\section{Two possible scenarios}
\label{scenarios}

In this section we will explore two different scenarios.

\subsection*{A. $\xi \ll 1$}

This case was considered in Ref.~\cite{GL}. Here the dilaton
expectation value $\nu$ is much larger than the present Planck mass,
see Eq.~(\ref{Planck}). For definiteness, we will choose a particular
value, $8\xi = 1/200$. From the required number of $e$-folds,
Eq.~(\ref{efolds}), this determines $\alpha$ to be of order one.

Let us study now the contribution of the different modes
to the CMB anisotropies. We first consider the continuum
spectrum of sub-curvature modes. The slow-roll parameters, given by 
Eqs.~(\ref{epet}) and (\ref{epet2}), become 
\begin{equation}
\label{eps1}
\epsilon \simeq {1\over200} \,, \hskip 1cm  \eta \simeq 0 \,,
\end{equation}
which determine the tilt of the primordial spectrum of density 
perturbations via Eq.~(\ref{tilt}) as
\begin{equation}
\label{tilt1}
n - 1 \simeq -6 \epsilon \simeq - 0.03 \,,
\end{equation}
which is compatible with large- and small-scale observations for $\Omega_0 
\gtrsim 0.4$, according to Ref.~\cite{WS}. The constraint on the 
amplitude of the angular power spectrum determines the value of $\lambda$ 
via Eq.~(\ref{constrC}) as
\begin{equation}
\label{lambda1}
\lambda \simeq 2.5 \times 10^{-5} {6\xi^3\over1+6\xi} \simeq 
        4 \times 10^{-14}\,.
\end{equation}
See Ref.~\cite{GL} for a range of values, under the assumption $\xi\ll1$.

We now consider the de Sitter vacuum super-curvature mode. This mode
exists, since $m_{{\rm F}}^2 < 2H_{{\rm F}}^2$ for $\xi\ll1$ from
Eqs.~(\ref{false}) and (\ref{false2}). The amplitude of curvature
perturbations that give rise to temperature anisotropies in the CMB is
constrained by Eq.~(\ref{condS}), which imposes the condition
Eq.~(\ref{constrS}). This is satisfied as long as $\alpha > 0.01$.
Since we are considering values of $\alpha \simeq 1$, the contribution
from the super-curvature mode will be negligible compared to that of
the sub-curvature modes, regardless of $\Omega_0$.\footnote{In the limit of 
small $\alpha$ and $\beta$, the $e$-foldings relation Eq.~(\ref{efolds}) 
gives $\alpha \simeq 10 \sqrt{\beta}$, so the super-curvature constraint 
will eventually become important once $\xi \lesssim 10^{-7}$.}

Consider now the bubble wall mode contribution to the CMB anisotropies. 
There are two possibilities, depending on the relative strength of the 
gravitational effects at tunneling \cite{bubble}. For $a^2b\ll1$ 
we are in the weak gravity regime of Ref.~\cite{LM}, and the condition on 
the parameters becomes, from Eq.~(\ref{constrW}),
\begin{equation}
\label{gamma1}
\gamma < 10^5 \,{1+6\xi\over6\xi}\,{M^3\over m_{{\rm Pl}}^3} \,.
\end{equation}
For $M \simeq 10^{-3} m_{{\rm Pl}}$ this gives $\gamma < 0.03$, which
is a reasonable bound on the coupling $\gamma$. On the other
hand, for $a^2b \gg 1$, condition Eq.~(\ref{condW}) is easily satisfied
for $\epsilon \simeq 1/200$, see Eq.~(\ref{wall2}).

\subsection*{B. $\xi\gg1$}

This case was considered in Ref.~\cite{Spokoiny}. The dilaton
expectation value $\nu$ is much smaller than the present Planck mass. 
For $\xi\gg1$, we have $\beta \simeq 4/3$ and the value 
of $\alpha$ is now determined from the required number of $e$-folds 
Eq.~(\ref{efolds}),
\begin{equation}
\label{xi2}
{4N\over3} \simeq \alpha - {4\over3} - \ln\Big({1+\alpha\over1+4/3}
        \Big)\,,
\end{equation}
which gives $\alpha \simeq 85$. This is a regime quite different from
the previous case.

We first consider the continuum spectrum of sub-curvature modes. The
slow-roll parameters are
\begin{equation}
\label{eps2}
\epsilon \simeq 1.85 \times 10^{-4} \,, \hskip 1cm
\eta \simeq - 0.016 \,,
\end{equation}
which determine the tilt of the primordial spectrum of density 
perturbations via Eq.~(\ref{tilt}) as 
\begin{equation}
\label{tilt2}
n - 1 \simeq 2\eta \simeq - 0.032\,.
\end{equation}
This is very similar to the previous case and thus compatible with
large- and small-scale observations for \mbox{$\Omega_0 \gtrsim 0.4$}. 
However, 
the constraint on the amplitude of the angular power
spectrum now determines the value of the combination $\lambda/\xi^2$ rather 
than $\lambda$ alone, as
\begin{equation}
\label{lambda2}
{\lambda\over\xi^2} \simeq 9 \times 10^{-10} \,.
\end{equation}
If we choose $\lambda \sim 1$, we have $\xi \sim 3 \times 10^4$,
which gives a very reasonable expectation value for 
the dilaton, from Eq.~(\ref{Planck}), of
\begin{equation}
\label{dilaton}
\nu \simeq 10^{-3} m_{{\rm Pl}} \simeq 10^{16} {\rm GeV}\,.
\end{equation}

We now consider the de Sitter vacuum super-curvature mode. This mode
also exists in this case, since $m_{{\rm F}}^2 < 2H_{{\rm F}}^2$ for
$\alpha\gg1$ from Eqs.~(\ref{false}) and (\ref{false2}). The amplitude of 
curvature perturbations is constrained by Eq.~(\ref{condS}), which imposes 
the condition Eq.~(\ref{constrS}). Since we have $\alpha\gg1$, the
contribution from this super-curvature mode will be negligible.

Concerning the bubble wall mode contribution, again there are two 
possibilities. In the weak gravity regime $a^2b \ll 1$, the condition on the 
parameters becomes
\begin{equation}
\label{gamma2}
\gamma < 10^5\,{M^3\over m_{{\rm Pl}}^3} \, \alpha^{5/2}\,.
\end{equation}
which gives a trivial constraint of $\gamma \lesssim 7$ for 
$M \simeq 10^{-3} m_{\rm Pl}$. For $a^2b\gg1$, the condition in 
Eq.~(\ref{condW}) is easily satisfied for $\epsilon \simeq 10^{-4}$.

\section{Matter era}
\label{matter}

One of the remaining issues is to make sure that after inflation the
scalar field $\varphi$ remains close to the minimum of its potential.
Deviations from this would result in time variations of the
gravitational constant, which are strongly constrained
\cite{willbook}.  During the radiation era the scalar field will
remain at the minimum due to the vanishing trace of the energy
momentum tensor, as seen from Eq.~(\ref{phieq}). However, during the
matter era the dilaton couples to the matter fluid and thus will be
subject to a force which shifts the field from its minimum.

However, it is easy to show that this effect is tiny. The relevant equation, 
from Eq.~(\ref{phieq}), is
\begin{equation}
\ddot\varphi+3H\dot\varphi+{\dot\varphi^2\over\varphi} =
        {4V(\varphi)-\varphi V'(\varphi) +\rho\over(1+6\xi)
        \varphi} \,.
\end{equation}
For a given $\rho$, there is a static solution at
\begin{equation}
\varphi_{{\rm st}}^2 = \nu^2 \left( 1+ \frac{2\rho}{\lambda \nu^4}
        \right) \,.
\end{equation}
Since the matter-era energy density is tiny in comparison to the 
inflationary energy density which determines $\lambda \nu^4$, the fractional 
shift in the gravitational constant at this static point is tiny, and so too 
is the energy density associated with the potential, which contributes only 
a minute fraction (perhaps $10^{-100}$!) of the critical density.

We have analyzed the detailed behavior, described in Appendix B. When 
matter domination starts, the field rises from its minimum to oscillate 
about the static point, which it does on a very rapid timescale. As $\rho$ 
decreases, the static point moves towards the true minimum (with the 
oscillation amplitude also decreasing though rather more slowly). At all 
times, the oscillations are of such small amplitude that general relativity 
holds to extremely high accuracy.

\section{Conclusions}

In this paper we have analyzed a variety of phenomenological
constraints on a recently proposed model of open inflation in the
context of induced gravity theories~\cite{GL}. The most stringent
constraints come from observations of the temperature anisotropies in
the microwave background. The model predicts a matter power spectrum
tilted to $n < 1$, which will be incompatible with observations if
the universe turns out to have $\Omega_0 \lesssim 0.4$. Otherwise, it is
possible to choose the parameters of the model so that it is in
agreement with observations.

During the matter era, the large dilaton mass and the extremely small
amplitude of oscillations around its vacuum expectation value ensure
that the theory approaches general relativity very efficiently,
passing all the post-Newtonian and oscillating gravitational coupling
tests.

{\em Final note:} We commented in the introduction that no method had been 
formulated to compute the gravitational wave spectrum, which we therefore 
did not consider. As we were revising for the final version of this paper, 
preprints appeared \cite{gravwaves} making significant progress in this 
direction. It will be interesting to apply these new results to specific 
open inflation models including the one discussed in this paper.

\section*{Acknowledgments}

J.G.B.~was supported in part by PPARC and A.R.L.~by the Royal Society.
We thank Anne Green, Andrei Linde and Martin White for useful discussions. 
J.G.B.~thanks Stanford University for its hospitality during part of this 
work, with the visit funded by NATO Collaborative Research Grant 
Ref.~CRG.950760. 
\appendix
\section{Open universe mode functions}

The open universe mode functions are discussed in Refs.\cite{LW,modes}

\subsection*{Sub-curvature modes}

The sub-curvature modes can be written as~\cite{Har,LW}
\begin{equation}
\label{PQL}
\Pi_{ql}(r) = N_{ql}\,\tilde\Pi_{ql}(r)\,,
\end{equation}
with
\begin{equation}
N_{ql} = \sqrt{\frac{2}{\pi}} \prod_{n=1}^l (n^2+q^2)^{-1/2} \,,
        \hspace{5mm} N_{q0} = \sqrt{\frac{2}{\pi}} \,,
\end{equation}
where the unnormalized modes $\tilde\Pi_{ql}(r)$ can be generated from the 
first two
\begin{eqnarray}
\label{modesC}
\tilde\Pi_{q0}(r) &=& {\sin qr\over\sinh r}\,,\\[1mm]
\tilde\Pi_{q1}(r) &=& {\coth r\,\sin qr - q\cos qr\over\sinh r}\,,
\end{eqnarray}
through the recurrence relation
\begin{eqnarray}
\label{recurC}
\tilde\Pi_{ql}(r) &=& (2l-1)\,\coth r \,\tilde\Pi_{q,l-1}(r) \nonumber
        \\[1mm]
&& \ -\, [(l-1)^2+q^2]\,\tilde\Pi_{q,l-2}(r)\,.
\end{eqnarray}

\subsection*{Super-curvature modes}

The first ($l\geq1$) multipoles are~\cite{GZ}
\begin{eqnarray}
\label{modesS}
\bar\Pi_{11}(r) &=& {1\over2}\Big[\coth r - {r\over\sinh^2 r}\Big]\,,\\[1mm]
\bar\Pi_{12}(r) &=& {1\over2}\Big[1 + {3(1-r \coth r)\over\sinh^2 r}\Big]\,.
\end{eqnarray}
The rest can be obtained with the recurrence relation
\begin{eqnarray}
\label{recurS}
\bar\Pi_{1l}(r) &=& {2l-1\over l-1}\,\coth r \,\bar\Pi_{1,l-1}(r) 
        \nonumber \\[1mm]
&& \ -\,{l\over l-1}\,\bar\Pi_{1,l-2}(r)\,.
\end{eqnarray}

\subsection*{Bubble wall modes}

The first ($l\geq2$) multipoles are~\cite{bubble}
\begin{eqnarray}
\label{modesW}
\bar\Pi_{22}(r) &=& {\sinh 4r - 8 \sinh 2r+12r \over 4\sinh^3r}\,,\\
\bar\Pi_{23}(r) &=& \\[1mm]
&&\hspace{-1cm} \,{\sinh 5r -15 \sinh 3r - 80 \sinh r +120 r \cosh r
        \over 8\sinh^4r}\,. \nonumber
\end{eqnarray}
The rest can be obtained from the recurrence relation
\begin{eqnarray}
\label{recurW}
\bar\Pi_{2l}(r) &=& {2l-1\over l-2}\,\coth r \,\bar\Pi_{2,l-1}(r) 
        \nonumber \\[1mm]
&& \ -\,{l+1\over l-2}\,\bar\Pi_{2,l-2}(r)\,.
\end{eqnarray}

\section{Matter era oscillations of the gravitational coupling}

Here we carry out a detailed analysis of the evolution of the dilaton
during the radiation and matter eras. Here we shall assume that the
oscillations are damped only by the Hubble expansion and not by any
particle decays -- if such decays were present the general
relativistic limit would be even more quickly approached.

The energy-momentum tensor conservation Eq.~(\ref{cons}) in the Jordan frame 
ensures $\rho a^{3(1+w)} =$ constant during the radiation $(w=1/3)$ and 
matter $(w=0)$ eras. In order to study the cosmological evolution during 
these eras, let us redefine our variables as
\begin{equation}
u = {\varphi^2\over\nu^2} - 1 \,, \hskip 1cm z = m t \,,
\end{equation}
where $m$ is given by
\begin{equation}
m^2 = {\lambda\nu^2\over1+6\xi} \,.
\end{equation}
The $\varphi$ equation of motion 
Eq.~(\ref{phieq}) and the Friedman equation become
\begin{eqnarray}
\label{ueq}
u'' + 3{a'\over a}\,u' + u & = & {2(\rho-3p)\over\lambda\nu^4}  \,, \\
\label{Friedman}
\left[2{a'\over a} + {u'\over1+u}\right]^2 & = & {1+6\xi\over6\xi} \times
        \nonumber \\
&&\hspace*{-1.5cm} \left[\left({u'\over1+u}\right)^2 + {u^2\over1+u} + 
        {8\rho\over\lambda\nu^4(1+u)} \right]\,,
\end{eqnarray}
where primes denote derivatives w.r.t. $z$. During the radiation era,
the right hand side of Eq.~(\ref{ueq}) vanishes and $u=u'=0$ is a stable
fixed point. Very soon one can neglect the $u$-terms in the Friedman
equation, and we find the radiation era attractor, $a'/a=1/2z$. The
scalar field equation of motion, $\,u'' + 3u'/2z + u = 0$, has an
exact solution,
\begin{equation}
z^{1/4}\,u(z) = c_1\,J_{1\over4}(z) + c_2\,Y_{1\over4}(z)\,,
\end{equation}
where $\{J, Y\}$ are Bessel functions. Its amplitude decays asymptotically 
as $u(z) \propto z^{-3/4}$, so we expect the matter era to start with 
initial conditions at $u=u'=0$.

During the matter era, $u=u'=0$ is a spiral attractor and 
we can always neglect the $u$-terms in the Friedman equation,
\begin{equation}
\left(\frac{a'}{a} \right)^2 \simeq \Big({1+6\xi\over6\xi}\Big) 
        {A\over a^3} = {4\over9z^2}\,,
\end{equation}
where $A = 2\rho a^3/\lambda\nu^4$ is a constant of order 
$10^{-120}$ in Planck units. The equation of motion for $u$ becomes  
\begin{equation}
u'' + {2\over z} u' + u = {A\over a^3} = {K\over z^2}\,,
\end{equation}
where $K=\beta/3$, see Eq.~(\ref{end}). There is an exact solution, 
\begin{equation}
z\,u(z) = c_1\,\sin z + c_2\,\cos z + K\,f(z)\,,
\end{equation}
where $f(z)$ is related to the Sine and Cosine Integral functions 
by~\cite{AS}
\begin{equation}
f(z) = {\rm Ci}(z)\,\sin z - {\rm Si}(z)\,\cos z =
\int_0^\infty {e^{-z\,t}\over1+t^2}\,dt\,.
\end{equation}
The late time $(z\to\infty$) behavior of $u$ is
$u(z) \propto \sin z/z$, with a large frequency of oscillations
\begin{equation}
\label{freq}
m = \Big[{\lambda\over8\pi\xi(1+6\xi)}\Big]^{1/2}
        \,m_{{\rm Pl}} \gg H_0\,,
\end{equation}
and an amplitude $|u'|\sim |u|\sim A/a_0^3$, which later decays as
$1/z$ at large $z$. The contribution of the scalar field to the total
energy density is therefore suppressed by an extra factor $A$ with
respect to the ordinary matter energy density, see
Eq.~(\ref{Friedman}). Since $A$ is so tiny, there are no constraints on the
parameters of the model from local experiments, see Ref.~\cite{Will},
and general relativity is a strong attractor of the equations of
motion.

Note that during the matter era the background dilaton field
oscillates very quickly, which might be thought could produce other
particles, like at the end of inflation. However, due to the extremely
small amplitude of oscillations, $|u|\sim A \sim 10^{-120}$, there is
no significant particle production and the field's energy can only
decay by redshifting away.


\end{document}